# Giant room-temperature nonlinearities from a monolayer Janus topological semiconductor

Jiaojian Shi[1,2,†], Haowei Xu[3,†], Christian Heide[4,5], Changan HuangFu[6], Chenyi Xia[1,2], Felipe de Quesada[1,2], Hongzhi Shen[8], Tianyi Zhang[9], Leo Yu[10], Amalya Johnson[1], Fang Liu[5,11], Enzheng Shi[8], Liying Jiao[6], Tony Heinz[5,10], Shambhu Ghimire[5], Ju Li[3,12], Jing Kong[9], Yunfan Guo[7*], Aaron M. Lindenberg[1,2,5*]

[1]Department of Materials Science and Engineering, Stanford University, Stanford, California 94305, United States

[2]Stanford Institute for Materials and Energy Sciences, SLAC National Accelerator Laboratory, Menlo Park, California 94025, United States

[3]Department of Nuclear Science and Engineering, Massachusetts Institute of Technology, Cambridge, Massachusetts 02139, United States

[4]Department of Applied Physics, Stanford University, Stanford, California 94305, United States

[5]Stanford PULSE Institute, SLAC National Accelerator Laboratory, Menlo Park, California 94025, United States

[6]Key Laboratory of Organic Optoelectronics and Molecular Engineering of the Ministry of Education, Department of Chemistry, Tsinghua University, Beijing 100083, China

[7]Key Laboratory of Excited-State Materials of Zhejiang Province, State Key Laboratory of Silicon Materials, Department of Chemistry, Zhejiang University, Hangzhou 310058, China

[8]School of Engineering, Westlake University, Hangzhou, 310024, China

[9]Department of Electrical Engineering and Computer Science, Massachusetts Institute of Technology, Cambridge, Massachusetts 02139, United States

[10]E. L. Ginzton Laboratory, Stanford University, Stanford, California 94305, United States

[11]Department of Chemistry, Stanford University, Stanford, California 94305, United States

[12]Department of Materials Science and Engineering, Massachusetts Institute of Technology, Cambridge, Massachusetts 02139, United States

*Corresponding author. E-mail: yunfanguo@zju.edu.cn (Y. G.); aaronl@stanford.edu (A. L.)

[†]These authors contributed equally to this work.

**Abstract:** Nonlinear optical materials possess wide applications, ranging from terahertz and mid-infrared detection to energy harvesting. Recently, the correlations between nonlinear optical responses and topological properties, such as Berry curvature and the quantum metric tensor, have stimulated great interest. Here, we report giant room-temperature nonlinearities in an emergent non-centrosymmetric two-dimensional topological material, the Janus transition metal dichalcogenides in the $1T'$ phase, which are synthesized by an advanced atomic-layer substitution method. High harmonic generation, terahertz emission spectroscopy, and second harmonic generation measurements consistently reveal orders-of-the-magnitude enhancement in terahertz-frequency nonlinearities of $1T'$ MoSSe (e.g., > 50 times higher than $2H$ MoS$_2$ for 18th order harmonic generation; > 20 times higher than $2H$ MoS$_2$ for terahertz emission). It is elucidated that such colossal nonlinear optical responses come from topological band mixing and strong inversion symmetry breaking due to the Janus structure. Our work defines general protocols for designing materials with large nonlinearities and preludes the applications of topological materials in optoelectronics down to the monolayer limit. This two-dimensional form of topological materials also constitute a unique platform for examining origin of the anomalous high-harmonic generation, with potential applications as building blocks for scalable attosecond sources.

Advances in nonlinear optics empower a plethora of applications, such as attosecond light sources based on high harmonic generation (HHG) and photodetectors for sensitive terahertz (THz) detection at elevated temperatures[1–4]. Inherently, the nonlinear optical properties of materials are connected with their magnetic structures[5,6], crystalline symmetries[7,8], and electronic band topologies. In particular, nontrivial band topologies lead to exotic electronic dynamics and enhanced optical responses[9–13]. Notable examples include anomalous HHG in various classes of topological materials[14–16]. The observation of enhanced optical responses in topological materials have been found primarily in three-dimensional systems until now[12–17]. Designing two-dimensional (2D) platforms with strong optical responses is advantageous for optoelectronic applications at the nanoscale with easy controllability and scalability, but by far is limited to topologically trivial materials such as graphene[18] and $2H$-phase transition metal dichalcogenides (TMDs)[19]. A promising topologically nontrivial candidate are the monolayer Janus TMDs (JTMDs) in the distorted octahedral ($1T'$) phase[3]. Similar to $1T'$ pristine TMDs[20–22], $1T'$ JTMDs are topologically nontrivial with an inverted bandgap in the THz regime (tens of meV). Generally, a topologically protected band structure and small electronic bandgap result in larger Berry connections, larger electronic interband transition rate, and thus stronger optical response. In addition, by replacing the top layer chalcogen atoms (e.g. sulfur) in the monolayer 1$T$'' TMDs with a different type of chalcogen (e.g. selenium), the resulting Janus structure leads to strong inversion asymmetry and electric polarization [23,24], which further improves the nonlinear optical response.

In this work, we report experimental observations of giant nonlinearities at THz frequencies in monolayer 1$T$'' JTMDs, which are synthesized via a room-temperature atomic-layer substitution (RT-ALS) method[25], under ambient conditions. It is revealed that, although the electromagnetic interaction occurs only in a single monolayer flake of 1$T$'' MoSSe (~ 10-20 μm in transverse size), the generation of mid-infrared high harmonics, THz emission, and infrared second harmonic generation are all exceptionally efficient. Further comparison with topologically trivial TMDs and theoretical analyses indicate that the key to such colossal THz-frequency nonlinearities is strong inversion symmetry breaking and topological band mixing. Our results suggest that 1$T$'' JTMDs is a promising material class that could lead to a new era in THz/infrared sensing using atomically-thin materials. Our results also deepen the understanding of the fundamental mechanisms underlying strong nonlinear optical responses, which could have a profound influence in, for example, room-temperature THz detection and clean energy harvesting.

**Multimodal nonlinearity characterization of 1$T$'' MoSSe**

The schematic illustration of multimodal characterization methods is shown in Fig. 1a. Our experiments investigated the THz-frequency nonlinearities of monolayer 1$T$'' MoSSe with three different techniques, i.e., high harmonic generation (HHG)[26,27], THz emission spectroscopy (TES), and second harmonic generation (SHG). These techniques access nonlinear coefficients with different orders (2nd to 18th order) and spectral ranges (THz to infrared). As a comparison, we also studied the responses of monolayer *2H* MoSSe, *1T'* $MoS_2$, and *2H* $MoS_2$ under the same measurement conditions. Such combined information unequivocally indicates giant THz-frequency nonlinearities for *1T'* MoSSe. As shown in Fig. 1b and 1c, the *1T'* phase of MoSSe and $MoS_2$ have distorted octahedral structures, with band inversion between metal *d*-orbitals and chalcogen *p*-orbitals[20]. In contrast, the *2H* phase is characterized by a trigonal prismatic structure and is topologically trivial. In this work, Janus *1T'* MoSSe and *2H* MoSSe (Fig. 1d) are both converted from their out-of-plane symmetric counterparts *1T'* $MoS_2$ and *2H* $MoS_2$ by room-temperature atomic-layer substitution method[2]. Highly reactive hydrogen radicals produced by a remote plasma were used to strip the top-layer sulfur atoms. Meanwhile, selenium vapor was supplied in the same low-pressure system to take the place of missing sulfur, resulting in the asymmetric Janus MoSSe in *1T'* phase and *2H* phase. To confirm the fidelity of material conversion, Raman scattering measurements were performed due to their sensitivity to the crystal lattice structure (Fig. 1e). For Janus *2H* MoSSe, the positions of the $A_{1g}$ mode (~288 $cm^{-1}$) and $E_{2g}$ mode (~355 $cm^{-1}$) are consistent with literature results[2]; For the *1T'* phase, the multiple *A'* modes of Janus *1T'* MoSSe located at ~226.2 $cm^{-1}$, ~298.4 $cm^{-1}$, ~429.8 $cm^{-1}$ agree well with the theoretical calculations, indicating the successful material substitution.

**High-harmonic generation efficiency boosted by topological bands and structural asymmetry**

We first show highly efficient HHG from a single monolayer flake of *1T'* MoSSe. The excitation source for HHG is mid-infrared (MIR) pulses with in-plane linear polarization at 5-µm wavelength, 1-kHz repetition rate, and ~ 20 MV/cm peak field strength (setup schematic shown in Supplementary Fig. 1). The HHG image acquired in *1T'* MoSSe (Fig. 2a) contains at least up to 18$^{th}$ order response, limited by our detection scheme. The even-order HHG, which is absent in bulk TMDs [27], is a direct consequence of the broken spatial symmetry of the monolayer systems. We varied the incident MIR polarization and observed nearly perfect cancellation of HHG intensity at specific angles to one of the crystallographic axes, indicating the HHG signal originates from a single flake instead of an average over many flakes with random orientations (shown in Fig. S4). This is consistent with the laser spot size (1/e$^2$ size) ~ 100 µm and the sparse flake-flake spacing (shown in Fig. S4). The HHG intensity of single flake 1$T'$ MoSSe are further compared with millimeter-scale 2$H$ MoS$_2$ under the same condition. Despite the irradiated flake being generally ~ 10 times smaller than the laser spot, the HHG of 1$T'$ MoSSe is over an order of magnitude stronger than that of the mm-size-flake 2$H$ MoS$_2$ with 100% coverage (Fig. 2b-d)[28]. The strong THz nonlinearity of 1$T'$ MoSSe is further confirmed by comparing it with other reference samples (2$H$ MoSSe and 1$T'$ MoS$_2$). Figure 2e shows the HHG spectrum of 2$H$ MoSSe, which has much weaker even-order harmonics than those of 1$T'$ MoSSe. Meanwhile, the 1$T'$ MoS$_2$ HHG, which is also topological nontrivial[20], shows relatively strong odd-order harmonics but no even-order harmonics, due to the inversion symmetry (Fig. 2f).

Although quantitative characterization of HHG efficiency is nontrivial, we find that the 1$T'$ MoSSe HHG efficiency at 18$^{th}$ order is 2-3 orders of magnitude higher than the representative HHG bulk solid ZnO - our data show similar cutoff orders and signal-to-noise ratios as the ZnO data under similar incident MIR fields[26]. Meanwhile, the 500-µm ZnO used[26] is theoretically shown to have an effective 100-200 nm thickness contributing to HHG generation[29], much thicker than monolayer 1$T'$ MoSSe (< 1 nm) with ~10 µm lateral sizes. Detailed analysis can be found in Supplementary Note 2. Further semi-quantitative HHG efficiency comparison with other literature[15,26,27,30–32] is summarized in Table 1 showing clear advantages of 1$T'$ MoSSe over most solid-state bulk or film samples. After normalizing with effective thickness, the efficiency is comparable with that found for HHG in liquid and gas samples [30,31] within an order of magnitude, despite requiring about two-orders-of-magnitude lower excitation fluence.

**Enhanced terahertz emission and second-harmonic generation efficiency**

Dramatic enhancements in the TES and SHG measurements further validate giant nonlinearities in 1*T'* MoSSe. Figure 3a shows the TES measurements under 800-nm laser excitation (details in Supplementary Fig. 2) on four kinds of CVD-grown samples (1*T'* MoSSe, 1*T'* MoS$_2$, 2*H* MoSSe, and 2*H* MoS$_2$), among which 1*T'* MoSSe shows distinctly higher THz emission efficiency. We do not observe a detectable signal in 1*T'* MoS$_2$ with the same excitation fluence, consistent with its centrosymmetric structure, which forbids second-order nonlinear response. The weak TES signal in 2*H* MoS$_2$ has been attributed to an inefficient surface photocurrent[33,34]. The augmented TES in 1*T'* MoSSe aligns with the theory that 1*T'* TMDs exhibit giant nonlinearities at THz frequencies[3]. The polarization analysis of the THz emission (Fig. 3b) reveals the emitted radiation is mainly polarized in the lab-frame *x*-direction and contains a slightly weaker *y*-direction component (axis definition shown in Fig. 3b). Based on our experimental configuration, the *x*-direction emission has contributions from both in-plane and out-of-plane photoresponses, while the *y*-direction emission originates only from in-plane photoresponses. Thus, our observation indicates a primary in-plane origin of TES signal but cannot rule out the out-of-plane contributions, consistent with theoretical predictions that in-plane photoresponses, including photocurrent and optical rectification effects, etc., are more substantial than the out-of-plane components[3]. For fixed excitation fluence, the peak THz field as a function of the pump polarization exhibits a sinusoidal modulation with a periodicity of approximately $\pi$ (Fig. 3c), reflecting the rank-two tensor nature of the photoresponses, which are second order in electric fields. Detailed analysis is included in Supplementary Fig. 5-10. Finally, the TES signal shows a linear dependence on the excitation fluence (Fig. 3d) at low fluences and saturates when the fluence exceeds 60 µJ/cm$^2$. Similar saturation behaviors have been observed in other TES measurements[33].

Figure 3e shows the SHG measurements on the four kinds of CVD-grown samples excited with 800-nm pulses (details in Supplementary Fig. 3). The SHG of several different flakes from each sample was measured to estimate the average intensity and flake-to-flake deviation. 2$H$ MoS$_2$, 1$T'$ MoSSe, and 2$H$ MoSSe show high SHG efficiency, and no SHG signal is detected in 1$T'$ MoS$_2$. In 1$T'$ MoSSe and 2$H$ MoSSe, SHG is further enhanced by a factor of 4 and 3 compared to monolayer 2$H$ MoS$_2$ respectively, for which high SHG efficiency has been extensively reported[19,35–37]. This highlights the importance of augmented inversion symmetry breaking in Janus structures, which improves even order nonlinearities. The SHG efficiency in Janus-type samples is further amplified in an angle-resolved SHG measurement (Fig. 3f) that is particularly sensitive to out-of-plane dipoles[23]. In this experiment, the incident angle of the 800-nm fundamental beam deviates from the normal incidence so that the tilted incident beam provides a vertical electric field and interacts with the out-of-plane dipoles effectively. To exclude other geometric factors, an $s$-polarized SHG $I_s$ induced by an in-plane dipole with the same collection efficiency is measured and used to normalize $p$-polarized SHG $I_p$ that contains out-of-plane dipole contribution at non-normal incidence. For 1$T'$ MoSSe and 2$H$ MoSSe, $I_p$/$I_s$ symmetrically increases as a function of the incident angle, while 2$H$ MoS$_2$ shows much smaller angle-dependent changes. This confirms the presence of out-of-plane dipoles in Janus-type samples.

**Theoretical origin of giant terahertz-frequency nonlinearity**

The experimental results above indicate that the optical nonlinearity of 1$T'$ MoSSe can be orders-of-magnitude (e.g., > 50 times higher for 18$^{th}$ order HHG; > 20 times higher for TES) stronger than those of 2$H$ MoSSe. To understand this effect, we examine the microscopic mechanism underlying the strong THz-frequency nonlinear responses in 1$T$' MoSSe. The band structures of 1$T$' MoSSe is shown in Fig. 4a. The band inversion of 1$T$' MoSSe happens around the Γ-point. Due to spin-orbit interaction, there is a band reopening at the ±Λ-points (marked in Fig. 4a). When the Fermi level is inside the bandgap, the interband transition dipole (Berry connection) $r_{mn}(\mathbf{k}) \equiv \langle m\mathbf{k}|r|n\mathbf{k}\rangle$ plays an essential role in optical processes[38], because it determines the strength of the dipole interaction between electrons and the electric fields. Here $r$ is the position operator, while $|m\mathbf{k}\rangle$ is the electron wavefunction at band $m$ and wavevector $\mathbf{k}$. In Fig. 4b-c, we plot $|r_{vc}(\mathbf{k})|$ of 2$H$ and 1$T$' MoSSe, where $v$ ($c$) denotes the highest valence (lowest conduction) band. For 2$H$ MoSSe, the maximum value of $|r_{vc}(\mathbf{k})|$ is around ~ 2 Å near the band-edge (±K points), while for 1$T$' MoSSe, $|r_{vc}(\mathbf{k})|$ can reach ~ 50 Å near the band-edge (±Λ points). Consequently, electrons in 1$T$' MoSSe would have stronger dipole interaction and hence faster interband transitions under light illumination. This is attributed to the topological enhancement – band inversions in topological 1$T$' MoSSe lead to wavefunction hybridization and hence larger wavefunction overlap between valence and conduction bands near the band edge, which accelerates the interband transitions[3,39,40]. The calculated first, second, and third-order nonlinear susceptibility of 2$H$ and 1$T$' MoSSe are shown in Figs. 4d-f. For $\omega \lesssim 0.5$ eV, the responses of 1$T$' MoSSe are significantly stronger than those of 2$H$ MoSSe. For $\omega \gtrsim 1$ eV, the responses of 1$T$' and 2$H$ MoSSe are close, consistent with experimental HHG, TES, and SHG measurements at different wavelengths. In the insets of Fig. 4d-f, we plot the $\mathbf{k}$-resolved contributions $I^{(i)}(\mathbf{k})$ to the optical susceptibility at $\omega = 1$ eV (see Methods section for details). Notably, the maximum value of $I^{(i)}(\mathbf{k})$ of 1$T$' MoSSe, located around the ±Λ points, is larger by orders-of-magnitude than that of 2$H$ MoSSe. This indicates that $\mathbf{k}$-points near the ±Λ points, which are influenced by topological enhancement, make major contributions to the total susceptibility even at $\omega = 1$ eV. Note that the bandgap of $1T'$ MoSSe is on the order of 10 meV, and thus interband transitions of electrons near the $\pm\Lambda$ points are far off-resonance with $\omega = 1$ eV photons. However, the contributions near the $\pm\Lambda$ points still dominate those at other $\boldsymbol{k}$-points where resonant interband transitions could happen. This again suggests the importance of the topological enhancement and the large interband transition dipoles near the $\pm\Lambda$ points. The topological enhancement gradually decays at large $\omega$. Consequently, the optical responses could be stronger in $2H$ MoSSe with $\omega \gtrsim 1$ eV.

The giant nonlinearities of $1T'$ JTMD, corroborated by both experimental and theoretical results above, support the colossal THz frequency photocurrent responses of $1T'$ JTMDs predicted by theory[3] and prelude that $1T'$ JTMD could serve as efficient THz detectors[10]. Our calculations indicate that the intrinsic photo-responsivity and noise equivalent power of the $1T'$ JTMD THz detector can outperform many current room-temperature THz sensors based on Schottky diodes or silicon field-effect transistors[41,42], albeit lower than the best pyroelectric detectors and bolometers[41] (Fig. 4g, see also Supplementary Note 1 and Supplementary Fig. 11). We foresee stacking multiple monolayer $1T'$ JTMDs and using field-enhancement structures[43] can further enhance the responsivity[44] and enable a facile usage of this detector for THz sensing purposes.

In conclusion, we demonstrate giant nonlinear responses in monolayer $1T'$ MoSSe, a prototype Janus topological semiconductor. Comparative experiments with different crystal phases ($2H$ vs. $1T'$) and symmetry types (Janus vs. non-Janus) indicate that $1T'$ MoSSe possesses orders-of-magnitude enhancement in HHG and second-order THz emission efficiency, and a few times enhancement in infrared SHG. Supported by theoretical calculations, our results elucidate that the remarkable enhancements originate from augmented structural asymmetry in Janus-type structures and topological band-mixing in 1$T'$ phases. The boosted HHG efficiency and the high fabrication versatility[25] of 1$T'$ JTMDs prelude a plethora of applications in attosecond physics, e.g., on-chip attosecond photonics[45,46] and light-wave electronics[47,48] in the monolayer limit. Meanwhile, the giant THz-frequency nonlinearities observed in this work could enable THz detection[49,50] with a large photo-responsivity at sub-A/W level and noise equivalent power down to the pW/$\sqrt{Hz}$ level.

**Table 1 | High-harmonic generation efficiency comparison of different materials.**

| Material | Phase | Thickness[i] (nm) | Pump field[ii] (V/Å) | Excitation λ (nm) | 17th order rate (unit)[iii] | 18th order rate (unit)[iv] | Reference |
|---|---|---|---|---|---|---|---|
| 1$T'$ MoSSe | Solid | ~1 | 0.2 | 5000 | ~ 100 (1) | ~20 (0.2) | This work |
| ZnO | Solid | ~$10^2$ | 0.2 | 3250 | < 0.1 ($10^{-3}$) | 0 (0) | Ref [26] |
| $Bi_2Se_3$ | Solid | ~1-10 | 0.25 | 5000 | < 0.01 ($10^{-4}$) | < 0.002 ($2*10^{-5}$) | Ref [32] |
| $MoS_2$ | Solid | ~1 | 0.4 | 5000 | < 1 ($10^{-2}$) | < 0.1 ($10^{-3}$) | Ref [27] |
| $CH_3OH$ | Liquid | ~$10^2$-$10^3$ | 1.5 | 4000 | 1 - 100 ($10^{-2}$ - 1) | 0 | Ref [31] |
| Xe | Gas | ~$10^3$-$10^7$ | 2.5 | 1850 | 0.1 - 1000 ($10^{-3}$ - 10) | 0 | Ref [30,33] |

Note: i) Effective thickness is estimated and adopted here. ii) The selected field strength of the excitation field. iii)-iv) The HHG efficiency is estimated and presented by emitted photon numbers / effective thickness / pump field. The numbers in the () are normalized to the 17$^{th}$ order for 1$T'$ MoSSe. Note the linear normalization to the pump field underestimates the difference between 1$T'$ MoSSe and other samples since HHG is a nonlinear process and incident field strength in 1$T'$ MoSSe is relatively low.

**Fig. 1 | Giant THz-frequency nonlinearities in 1$T'$ MoSSe. a**, Schematic illustration of THz emission spectroscopy (TES), mid-infrared high-harmonic generation (HHG), and near-infrared second harmonic generation (SHG) in 1$T'$ MoSSe. **b**, The lattice structure of 1$T'$ MoSSe, 1$T'$ $MoS_2$, 2$H$ MoSSe, 2$H$ $MoS_2$. **c**, Schematic illustration of the topological band inversion in the 1$T'$ phase (left) compared with the 2$H$ phase. The colormap indicates the wavefunction contributed by different electron orbitals, and the 1$T'$ phase exhibits a hybridization between the valence and the conduction bands. **d**, Optical images of 1$T'$ MoSSe (top) and 2$H$ MoSSe (bottom). **e**, Experimental and theoretical Raman spectrum of 1$T'$ MoSSe and experimental Raman spectrum of 2$H$ MoSSe.

**Fig. 2 | Extremely efficient mid-infrared high harmonic generation (HHG) in 1$T'$ MoSSe. a,** HHG images of 1$T'$ MoSSe observed by CCD camera. HHG extends up to ~ 5 eV and is limited mainly by the cutoff of detection optics (e.g., aluminum mirrors and grating). **b**, HHG spectrum of 1$T'$ MoSSe shown in the blue curve is over an order of magnitude stronger than the HHG from macroscopic monolayer 2$H$ $MoS_2$ shown as the red dashed line. The inset shows 1$T'$ MoSSe HHG intensity as a function of MIR incident polarization angles. The cancellation of a few orders at some polarization angles indicates the signal is generated from a single flake. **c-d**, Quantitative comparison of HHG intensity between 1$T'$ MoSSe and 2$H$

MoS$_2$. Although the coverage of the 1$T$' MoSSe sample is ~ 10 times lower than wafer-scale 2$H$ MoS$_2$, the HHG signal is enhanced by over an order of magnitude, and orders as high as 18 or more are observed in 1$T$' MoSSe while only up to 15$^{th}$ order can be observed in macroscopic monolayer 2$H$ MoS$_2$. **e**, The HHG spectrum of 2$H$ MoSSe taken under the same conditions. **f**, HHG spectrum of 1$T$' MoS$_2$. Even orders of 2$H$ MoSSe and 1$T$' MoS$_2$ are weak and indetectable compared with 1$T$' MoSSe.

**Fig. 3 | THz emission spectroscopy and second harmonic generation of 1$T$' MoSSe, 1$T$' MoS$_2$, 2$H$ MoSSe and 2$H$ MoS$_2$. a,** Left plot shows a schematic of THz emission setup. 1$T$' MoSSe shows a dramatically enhanced THz emission signal compared with three other types of monolayer TMDs. **b**, Left plot shows a schematic of polarization analysis of THz emission, which shows the emission contains a major in-plane component with a possible out-of-plane contribution. **c**, Dependence of peak THz field on excitation polarization. a.u., arbitrary units. **d**, Scaling of THz emission shows a linear dependence to incident power at low fluences and saturation at higher fluences. **e**, Left plot shows a schematic of second harmonic generation with normally incident 800-nm excitation. The right plot shows SHG intensity statistics of five different flakes in each sample and shows SHG is enhanced in 1$T$' MoSSe and 2$H$ MoSSe compared with 1$T$' MoS$_2$ and 2$H$ MoS$_2$. **f**, The left plot shows the schematic of the angle-resolved SHG setup that measures out-of-plane dipole. The beam position at the objective back aperture is scanned perpendicular to the incident beam direction with a motorized stage, which tunes the incident angle accordingly. The right plot shows the angle-dependent SHG intensity ratio between $p$ and $s$ polarization ($I_p$ and $I_s$) in 1$T$' MoSSe, 2$H$ MoSSe, and 2$H$ MoS$_2$. In the 1$T$' MoSSe and 2$H$ MoSSe, the $I_p/I_s$ ratio increases at non-normal incidence angles, indicative of out-of-plane dipoles. In 2$H$ MoS$_2$, almost no change is observed as the incident angle varies.

**Fig. 4 | Theoretical calculations of the nonlinear optical response of 2$H$ and 1$T$' MoSSe. a,** Band structure of 1$T$' MoSSe. The energy is offset to the valence band maximum. The band edge of 1$T$' MoSSe is located at the $\Lambda$ point, which is marked in **a**. **b-c**, The magnitude of the interband transition dipole $|r_{vc}(\mathbf{k})|$ for (**b**) 2$H$ and (**c**) 1$T$' MoSSe in the first Brillouin zone. The colormap is in the unit of Å. **d-f**, First, second, and third-order optical responses of 1$T$' and 2$H$ MoSSe as a function of incident light frequency $\omega$. Insets of (**d-f**) show the $\mathbf{k}$-resolved contributions (with arbitrary units) to the total response function at $\omega = 1$ eV for (left) 2$H$ and (right) 1$T$' MoSSe. Red (blue) arrows in (**b-c**) and insets of (**d-f**) denote the $\pm K$ ($\pm \Lambda$) points, which are the bandedge of 2$H$ (1$T$') MoSSe. In insets of (**d-f**), the Brillouin zone is zoomed in around the $\pm \Lambda$ points for 1$T$' MoSSe to give better visibility. **g**, The NEP (left $y$-axis) and photo-responsivity (right $y$-axis) of 1$T$' JTMD THz detector as a function of the light frequency $\omega$.

**Acknowledgments:** J.S., C.X., F.Q., A.L. acknowledges the support from the Department of Energy, Office of Basic Energy Sciences, Division of Materials Sciences and Engineering, under contract DE-AC02-76SF00515. Y.G. acknowledges the financial support from Zhejiang University.  H. X. and J. L. were supported by an Office of Naval Research MURI through grant #N00014-17-1-2661. E.S. and H.S. acknowledge the financial support from Research Center for Industries of the Future at Westlake University, National Natural Science Foundation of China (grant no.52272164). J.K. and T.Z. acknowledges the financial support from US Department of Energy (DOE), Office of Science, Basic Energy Sciences under Award DE‐SC0020042.

**Author contributions:** J.S., H.X., Y.G. and A.L. designed the study; Y.G. performed the Janus material synthesis and Raman characterization; H.X. and J.L. performed the theoretical analyses and ab initio calculations.; C.H. and J.S. performed HHG measurements under the supervision of S.G. and A.L.; C.X. and J.S. performed TES measurements under the supervision of A.L.; J.S.; C.H.F. synthesized typical transition metal dichalcogenides under the supervision of L. J. F.Q. and L.Y. performed SHG measurements under the supervision of A.L. and T.H.; A.J. synthesized wafer-scale $2H$ $MoS_2$ under the supervision of F.L.; H.S., T.Z, E.S., J.K. participated in data analysis; J.S., H.X., Y.G. wrote the manuscript; All authors read and revised the manuscript.

**Additional information:** Authors declare no competing interests. All data are available in the manuscript or supplementary material. All materials are available upon request to Y.G. and A.L.

**Competing financial interests:**

The authors declare no competing financial interests.

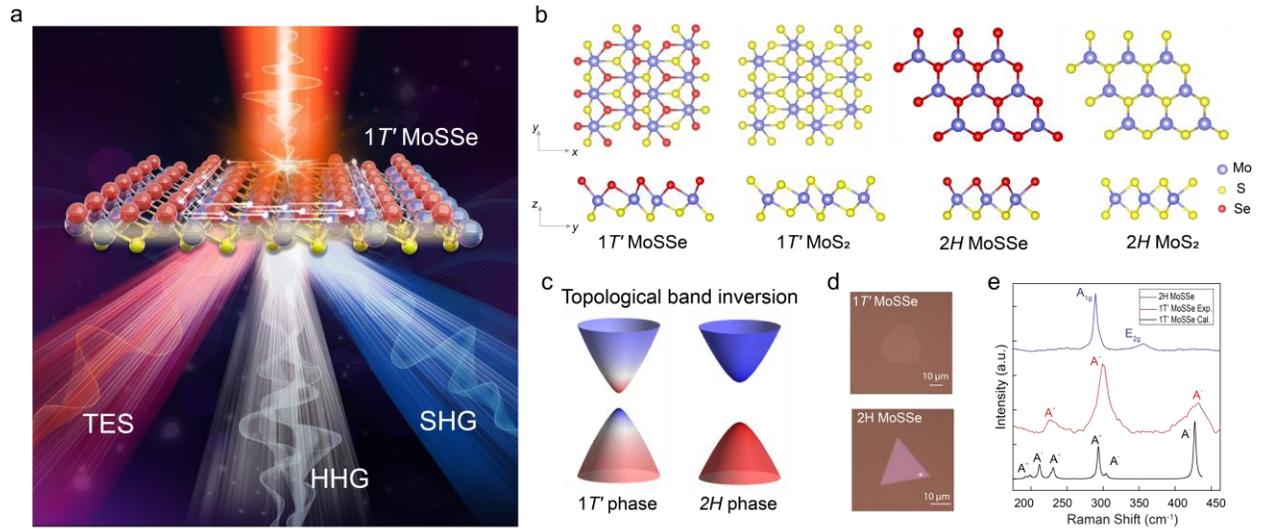

**Fig. 1| Giant THz-frequency nonlinearities in 1$T'$ MoSSe. a**, Schematic illustration of THz emission spectroscopy (TES), mid-infrared high-harmonic generation (HHG), and near-infrared second harmonic generation (SHG) in 1$T'$ MoSSe. **b**, The lattice structure of 1$T'$ MoSSe, 1$T'$ MoS$_2$, 2$H$ MoSSe, 2$H$ MoS$_2$. **c**, Schematic illustration of the topological band inversion in the 1$T'$ phase (left) compared with the 2$H$ phase. The colormap indicates the wavefunction contributed by different electron orbitals, and the 1$T'$ phase exhibits a hybridization between the valence and the conduction bands. **d**, Optical images of 1$T'$ MoSSe (top) and 2$H$ MoSSe (bottom). **e**, Experimental and theoretical Raman spectrum of 1$T'$ MoSSe and experimental Raman spectrum of 2$H$ MoSSe.

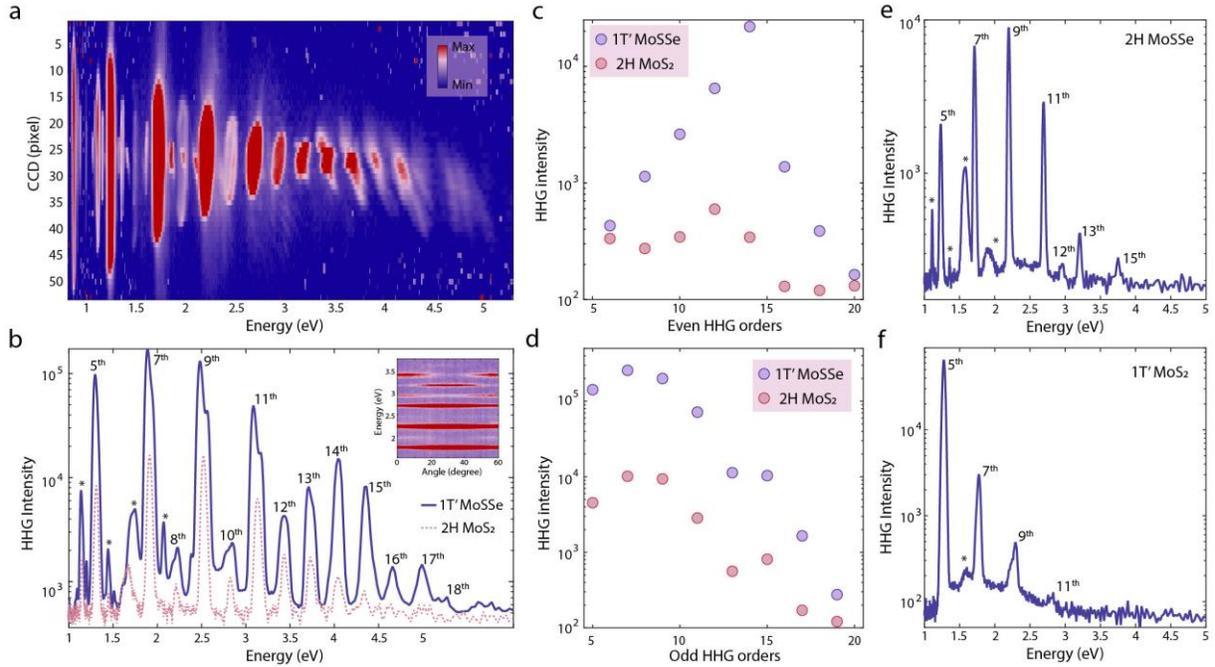

**Fig. 2| Extremely efficient mid-infrared high harmonic generation (HHG) in 1*T'* MoSSe. a,** HHG images of 1*T'* MoSSe observed by CCD camera. HHG extends up to ~ 5 eV and is limited mainly by the cutoff of detection optics (e.g., aluminum mirrors and grating). **b**, HHG spectrum of 1*T'* MoSSe shown in the blue curve is over an order of magnitude stronger than the HHG from macroscopic monolayer 2*H* $MoS_2$ shown as the red dashed line. The inset shows 1*T'* MoSSe HHG intensity as a function of MIR incident polarization angles. The cancellation of a few orders at some polarization angles indicates the signal is generated from a single flake. **c-d**, Quantitative comparison of HHG intensity between 1*T'* MoSSe and 2*H* $MoS_2$. Although the coverage of the 1*T'* MoSSe sample is ~ 10 times lower than wafer-scale 2*H* $MoS_2$, the HHG signal is enhanced by over an order of magnitude, and orders as high as 18 or more are observed in 1*T'* MoSSe while only up to $15^{th}$ order can be observed in macroscopic monolayer 2*H* $MoS_2$. **e**, The HHG spectrum of 2*H* MoSSe taken under the same conditions. **f**, HHG spectrum of 1*T'* $MoS_2$. Even orders of 2*H* MoSSe and 1*T'* $MoS_2$ are weak and indetectable compared with 1*T'* MoSSe.

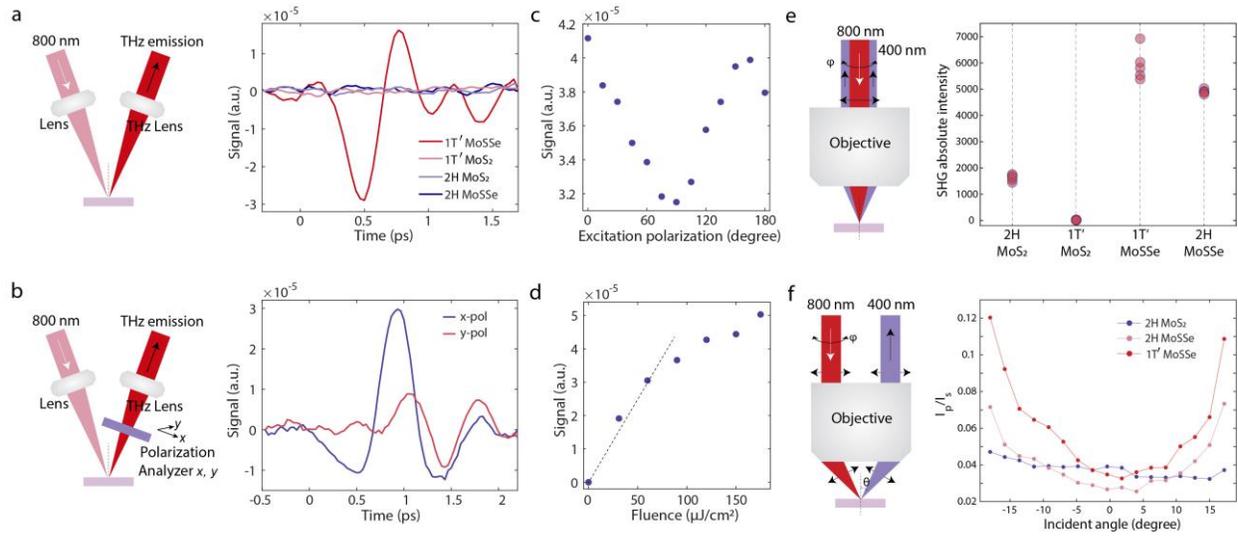

**Fig. 3| THz emission spectroscopy and second harmonic generation of 1$T'$ MoSSe, 1$T'$ MoS$_2$, 2$H$ MoSSe and 2$H$ MoS$_2$. a,** Left plot shows a schematic of THz emission setup. 1$T'$ MoSSe shows a dramatically enhanced THz emission signal compared with three other types of monolayer TMDs. **b**, Left plot shows a schematic of polarization analysis of THz emission, which shows the emission contains a major in-plane component with a possible out-of-plane contribution. **c**, Dependence of peak THz field on excitation polarization. a.u., arbitrary units. **d**, Scaling of THz emission shows a linear dependence to incident power at low fluences and saturation at higher fluences. **e**, Left plot shows a schematic of second harmonic generation with normally incident 800-nm excitation. The right plot shows SHG intensity statistics of five different flakes in each sample and shows SHG is enhanced in 1$T'$ MoSSe and 2$H$ MoSSe compared with 1$T'$ MoS$_2$ and 2$H$ MoS$_2$. **f**, The left plot shows the schematic of the angle-resolved SHG setup that measures out-of-plane dipole. The beam position at the objective back aperture is scanned perpendicular to the incident beam direction with a motorized stage, which tunes the incident angle accordingly. The right plot shows the angle-dependent SHG intensity ratio between $p$ and $s$ polarization ($I_p$ and $I_s$) in 1$T'$ MoSSe, 2$H$ MoSSe, and 2$H$ MoS$_2$. In the 1$T'$ MoSSe and 2$H$ MoSSe, the $I_p/I_s$ ratio increases at non-normal incidence angles, indicative of out-of-plane dipoles. In 2$H$ MoS$_2$, almost no change is observed as the incident angle varies.

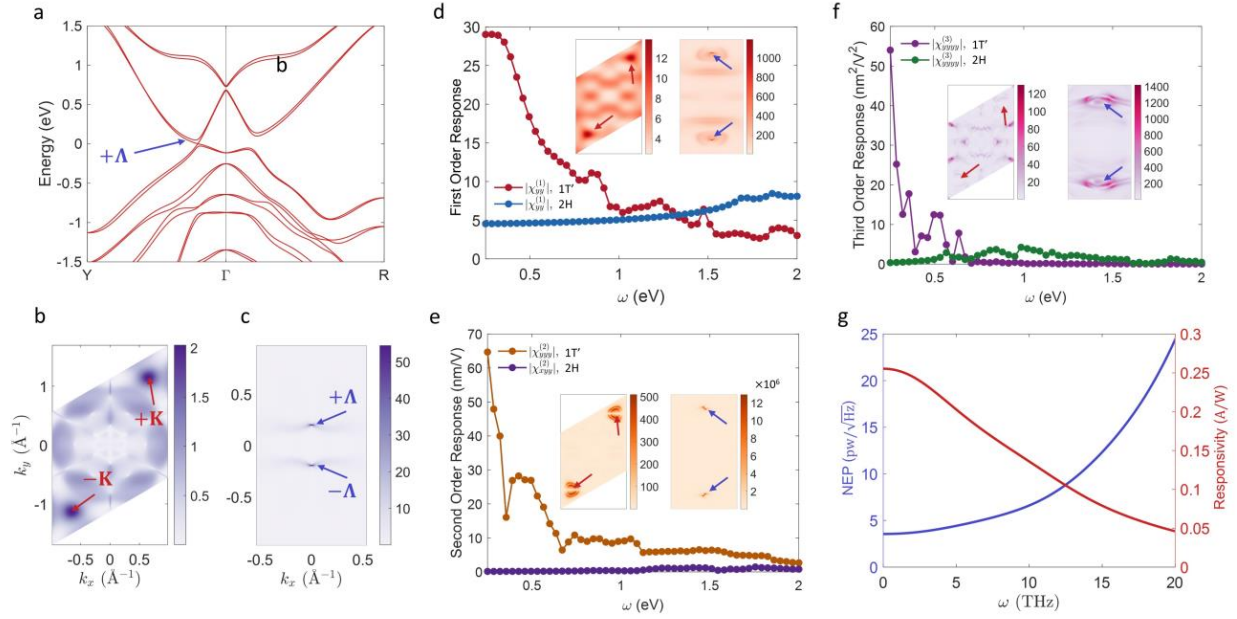

**Fig. 4| Theoretical calculations of the nonlinear optical response of 2H and 1T' MoSSe. a,** Band structure of 1T' MoSSe. The energy is offset to the valence band maximum. The band edge of 1T' MoSSe is located at the $\Lambda$ point, which is marked in **a**. **b-c**, The magnitude of the interband transition dipole $|r_{vc}(\mathbf{k})|$ for (**b**) 2H and (**c**) 1T' MoSSe in the first Brillouin zone. The colormap is in the unit of Å. **d-f**, First, second, and third-order optical responses of 1T' and 2H MoSSe as a function of incident light frequency $\omega$. Insets of (**d-f**) show the $\mathbf{k}$-resolved contributions (with arbitrary units) to the total response function at $\omega = 1$ eV for (left) 2H and (right) 1T' MoSSe. Red (blue) arrows in (**b-c**) and insets of (**d-f**) denote the $\pm K$ ($\pm\Lambda$) points, which are the bandedge of 2H (1T') MoSSe. In insets of (**d-f**), the Brillouin zone is zoomed in around the $\pm\Lambda$ points for 1T' MoSSe to give better visibility. **g**, The NEP (left *y*-axis) and photo-responsivity (right *y*-axis) of 1T' JTMD THz detector as a function of the light frequency $\omega$.